# Long-term secure distributed storage using quantum key distribution network with third-party verification

Mikio Fujiwara[1], Ryo Nojima[1], Toyohiro Tsurumaru[2], Shiho Moriai[1], Masahiro Takeoka[1], and Masahide Sasaki[1]

[1]National Institute of Information and Communications Technology, Nukui-kita, Koganei, Tokyo 184-8795 Japan.
[2]Mitsubishi Electric Corp., Kanagawa, 247-8501 Japan.

Corresponding author: Mikio Fujiwara (email: fujiwara@ nict.go.jp).

This work was partly supported by Council for Science, Technology and Innovation (CSTI), Cross-ministerial Strategic Innovation Promotion Program (SIP), "Photonics and Quantum Technology for Society 5.0" (Funding agency: QST) and JSPS KAKENHI Grant Number JP18H05237.

**ABSTRACT** The quantum key distribution (QKD) network [1,2] with Vernam's One Time Pad (OTP) encryption [3] and secret sharing [4] are powerful security tools to realize an information theoretically secure (ITS) distributed storage system. In [5], a single-password-authenticated secret sharing (SPSS) scheme based on the QKD network and Shamir's secret sharing [4] was experimentally demonstrated; it confirmed ITS data transmission, storage, authentication, and integrity. To achieve data integrity, an ITS message authentication code (MAC) tag is employed and a data owner of the secret sharing performs both the MAC tag generation and verification. However, for a scenario in which the data owner and end users are different entities, the above approach may not work since the data owner can cheat the end users. In this paper, we resolve this problem by proposing an ITS integrity protection scheme employing a third-party verification with time-stamp. The ITS integrity protection is realized by two steps: integrity check by the data owner at data reconstruction, and data integrity certification by the data owner, the end user, and the third-party verifier using a MAC based on universal$_2$ hash function [6] and random number provided from the QKD network [7]. In addition to introduce the third-party verifier, we institute "a trusted calculator" which computes shares of the data and MAC tags and sends MAC tags to the third-party verifier. The random number used in calculating MAC tag is stored in the trusted calculator. We implement this scheme on the SPSS system installed in the Tokyo QKD Network [8] and evaluate the performance of the third-party verification in view of attack scenarios on this system. In addition, we demonstrate a simple share renewal function based on verifiable secret sharing scheme which ensures the data integrity for a certain practical period based on the hardness of discrete logarithm problem. To our best knowledge, this is the first demonstration of ITS secure data transmission, storage, authentication, and data integrity with the third-party in a metropolitan area network.

**INDEX TERMS** Quantum key distribution, Tokyo QKD Network, secret sharing, time-stamp, third-party verification

## I. INTRODUCTION

Long-term protection of integrity, authenticity, and confidentiality are required for critical information assets, for example, medical information such as genomic data and classified national information. Information leakage may cause criminal and/or civil penalties to data owners and system providers [9, 10]. In addition, these information assets should be available against disasters and various technical faults, in view of reasonable "Business Continuity Plan (BCP)s".

Distributed data backup in distant places is one of effective solutions to the BCP issue. To combine this approach with the long-term security, transmission and storage of critical data must be strictly protected even against future technologies including quantum computers, which will be realistic, taking into account recent efforts on its development around the world [11,12].

In this regard, a distributed data storage system with information theoretical security, which consists of quantum key distribution (QKD) [1,2] links and a secret sharing protocol [4], is a very promising solution. A QKD link enables two remote users to share Information Theoretically Secure (ITS) key (random number). Vernam's One Time Pad (OTP) [3] with such keys provides ITS data transmission.

Shamir's secret sharing scheme [4] can realize ITS storage system, if data transmission and authentication is carried out in an ITS way.

These two schemes can provide ITS confidentiality of transmission and storage of data. Besides this, integrity protection of data is also an important concern in practice. Here the integrity means that illegitimate or accidental changes of data can be discovered (see, e.g., Ref. [13]). And a correct data must be shared between the data owner and the ultimate user of the data (we name him as an end user). This is the main topic of this paper.

In our previous work [5], we have proposed an ITS distributed storage system with ITS authentication based on a user-friendly single-password-authenticated secret sharing (SPSS) scheme and secure transmission using QKD and demonstrated it in the Tokyo QKD network [8]. In this scheme, the ITS message authentication code (MAC) tag, generated by the universal$_2$ hash function [6,14,15] with a password used as the key, is added to the storage data. Thereby, the data owner can confirm the ITS integrity of the data by oneself when the data owner reconstructs the data.

We note that several methods have been reported to also protect data integrity against share holders' cheating by using hash functions [16,17], while our method [5] enables the data owner to know falsification of the data at a single-password authentication simultaneously.

(Outline of this scheme is summarized in section 3.)

In some cases, it is valuable that the function of the integrity protection be provided by a third-party. If the data owner and the end user are different, for example, saving a testament, the integrity check of the testament data by the third-party is extremely important since either the data owner or the end user may tamper with the data. To provide the third-party integrity check, in our subsequent work, we proposed Long-term INtegrity and COnfidentiality protection System (LINCOS) under some security assumptions [13]. In this system, the secret data is reconstructed at regular intervals, and integrity of the data is protected by commitment renewal guaranteed by the evidence service and time-stamp service (i.e. the third-party) in the authenticated network. Such a verification system with the third party is widely used in time-stamp service [18]. It is known that commitment schemes used for data integrity cannot achieve ITS binding and ITS hiding, simultaneously [19]. In [13], therefore, only ITS hiding is employed.

In this paper, we propose another third-party verification scheme with MAC, which realizes both the ITS binding and hiding simultaneously. The proposed system is based on the distributed storage system in [5] but additionally introduces the third-party verifier. We use universal$_2$ hash function [6,14,15] to calculate MAC tags, which preserves the ITS confidentiality. Unlike in the case of [5], where the MAC tag-generation and verification are both performed by the data owner alone, in the present scheme, the MAC tag-generator and the MAC verifier are different persons. A problem of this setting is that the MAC generator enables to falsify the MAC tags easily due to the property of the universal$_2$ hash function. Therefore, in the proposed scheme, we introduce "a trusted calculator" [20] for the MAC tag-generator. Such a concept of secure computation using a trusted hardware, which is trusted but has small long-term memory capacities, is used in practices (an example of such a device is in [21,22]). In our system, the trusted calculator computes shares and MAC tags. In calculation of MAC tags, the calculator uses random number provided from a QKD network and memorizes this random number until verification by the end user.

The proposed scheme is experimentally demonstrated in a QKD network testbed, referred to as the Tokyo QKD Network [8]. A distributed storage system with four share holders is implemented, which is supported by the five-nodes QKD network. In the same system, we also implement a simple share renewal function based on Pedersen's verifiable secret sharing scheme, which may be of independent interest on practical implementation of the share renewal function.

The paper is organized as follows. In section 2, we describe our verification scheme with the trusted calculator which calculates shares and MAC tags, and memorizes random numbers used in tag-generation of the MAC. And summary of SPSS scheme is described in section 3. The experimental results of the verification scheme with the simple share renewal function are given in section 4. Conclusion is summarized in section 5.

## II. THIRD-PARTY VERIFICATON SCHEME
### 2.1 Motivation and basic ideas

Conventional secret sharing schemes usually consist only of a data owner and share holders. We consider here a new setting where there is an additional player, called an end user, who receives the data from the data owner and casts doubt on its integrity.

The goal of our third-party verification function is to resolve a dispute about data integrity, even if a malicious end user entertains false doubts about the secret data, disclosed by the data owner. Figure 1 shows a basic configuration of a secret sharing scheme with data integrity.

Note here that a third-party, the verifier, is newly introduced. This is necessary in order to ensure data integrity to an end user [13]. As mentioned in introduction, the efficient way to achieve data integrity is to use commitment scheme [13]. However, in commitment schemes, ITS binding and ITS hiding have not been achieved simultaneously [19]. As a solution to break through this limitation, we adopt a trusted server with small long-term memory as a key player to realize ITS binding and ITS hiding in a distributed data backup system on the QKD network. The overview of our scheme is described below.

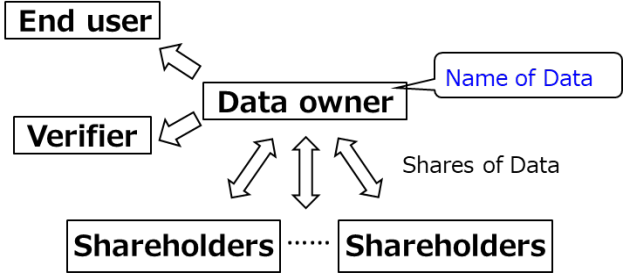

**FIGURE 1.** Conceptual view of a secret sharing scheme with data integrity. The verifier enforces the third-party verification.

## 2.2 Setting and the goals

Given a $t_{SH}$-out-of-$n_{SH}$ secret sharing (SS) scheme $SS$, we add to it a third-party verification function as follows. The underlying SS scheme $SS$ can be arbitrary, hence we do not specify its details here.

### 2.2.1 Players and their roles:
We consider the situation where there are the following players with the following roles (Fig. 2).

- **Data owner** is the original owner of data $D \in \mathcal{D}$ ($\mathcal{D}$ is the set of all possible data). He asks the share calculator to register and store $D$. Whenever requested by an end user, he must retrieve and release $D$.
- **Share calculator** calculates shares $s = (s_1, \ldots, s_{n_{SH}})$ of $D$ using the SS scheme $SS$, and sends $s_h$ to share holder $h \in \{1, \ldots, n_{SH}\}$.
  He also generates the MAC tag $\sigma$ using random number $R_{MAC}$ provided from the QKD network. The MAC tag $\sigma$ serves as the evidence that $D$ was received at time $t_1$. He then asks the verifier to store $\sigma$. He memorizes $R_{MAC}$.
- **Share holders.** There are $n_{SH}$ share holders. Each of them, indexed by $h \in \{1, \ldots, n_{SH}\}$, stores a share $s_h$ of $D$ calculated by the share calculator.
- **Verifier** receives and stores tag $\sigma$ delivered from the share calculator, along with the time $t_2$ of its receipt.
- **End user:** An end user requests data owner to send $D$ and $t_1$. He can detect a possibility of receiving false data with the help of the share calculator and the verifier. We stress that the end user may not be specified until the date reconstruction phase.

As mentioned above, we assume that the share calculator is trusted but has small long-term memory capacities. That is, the share calculator can store small data (the random numbers) for a long time but can store large data only for short time (the original data and its shares) [21,22]. In other words, the share calculator is fully trusted but kept minimal for a practical purpose. One of important roles of the share calculator is to store the random numbers used in calculation of MAC tags. This makes it impossible for the data owner to guess the MAC tag of the data.

### 2.2.2 Goals (Security Criteria)

We let $k \in \mathbb{N}$ be the security parameter. Typically, $k$ is 256. Our goal in the scheme defined below is to fulfill the following security criteria. Below, $|\mathcal{D}|$ denotes the cardinality of the set $\mathcal{D}$ of all possible data $D$.

- **SC1 (Integrity from the viewpoint of the end user):** Except with a probability $\leq 2^{-k} \log_2 |\mathcal{D}|$, an honest end user can detect when the data owner reveals the data $D''$, which differs from $D$ that was registered.
- **SC2 (Integrity from the viewpoint of the data owner):** Except with a probability $\leq 2^{-k} \log_2 |\mathcal{D}|$, the data owner, if honest, can refute a false claim made by an end user that received a data $D''$, which differs from $D$.
- **SC3 (Secrecy):** The amount of information that the end user or the verifier obtains concerning $D$, prior to the reveal phase, is less than $k$ bits.

## 2.3 Description of our scheme

In this subsection we defined our scheme.

### 2.3.1 Assumptions

We begin by the listing the underlying assumptions.

- **A1 (Share calculator):** The share calculator is trusted, meaning that he follows the procedure specified in the next subsection correctly and leaks no information.
- **A2 (Data reconstruction phase):** At the data reconstruction phase of $SS$, the share calculator can verify that he indeed recovered the correct data $D$.
- **A3 (Verifier):** The verifier is honest but curious, meaning he follows the procedure specified in the next subsection correctly but may leak information.
- **A4 (Channels with the perfect security in the ITS sense):** Each pair of players (all players listed in Sec. 2.1.1) are connected by a channel with the perfect secrecy and the perfect authenticity in the ITS sense. That is, every player pair can use a channel where no eavesdropping or modification is possible, even when equipped with an unlimited computation power.

Several remarks are in order concerning these assumptions.
First, none of assumptions above restricts behaviors of the data owner and the end user; hence these two players can always deviate from our scheme in any way.
Second, item A2 can be guaranteed e.g. by assuming (i) that share calculators always submit correct shares, or (ii) that the underlying SS schemes $SS$ is equipped with certain cheater detection mechanisms (see e.g. Refs. [5,16,17]).
Third, item A3 entails that the verifier, when asked, answers the correct values of $\sigma$ and $t_2$, but does not necessarily keep them secret.
Finally, we stress that item A4 (the perfectly secure channels) can easily be realized in QKD networks, where every player pair $p, q$ has access to an arbitrarily long secret key $k_{pq}$ with the perfect security. The secrecy of the channel can be guaranteed by OTP using $k_{pq}$. The authenticity can be guaranteed, e.g., by message authentication codes with the ITS, which uses $k_{pq}$ only once (see e.g. Ref. [15]).

### 2.3.2 Specification of the MAC tag σ

In generating, the MAC tag σ, the share calculator uses an almost universal$_2$ hash function $u$, which has the following property.

**Lemma 1** (Existence of an almost universal$_2$ function): There exists a function $u: \mathcal{R}_{MAC} \times \mathcal{D} \to \{0,1\}^k$ for which

$$|\mathcal{R}_{MAC}|^{-1} \sum_{R_{MAC} \in \mathcal{R}_{MAC}} 1[u(R_{MAC}, a) = u(R_{MAC}, a')] \leq 2^{-k} \log_2 |\mathcal{D}|$$

holds for $\forall a, a' \in \{0,1\}^*$ satisfying $a \neq a'$. Here $|A|$ denotes the cardinality of set $A$, and $1[Q]$ is the function that equals 1 if proposition $Q$ holds, and 0 otherwise.

*Proof*: Let $\mathcal{R}_{MAC}$ be a finite field $\mathbb{F}_q$ with the size $q$ satisfying $2^k \geq q \geq k^{-1} 2^k$. Then let $u$ be the hash function family given in Theorem 3.5 of Ref. [23].

In order to guarantee the ITS, we require that variable $R_{MAC}$ be a true random number [15]; e.g. a quantum random number provided from a QKD system. We also require that $R_{MAC}$ be generated newly every time the scheme is started [15].

### 2.3.3 Procedures

The procedure of our scheme consists of two phases, the data registration phase and the data reconstruction phase. The latter includes the integrity check of the reconstructed data $D'$ corresponding to $D$.

**(1) Data registration phase:**
1.1 **Initiation by the data owner:** The data owner sends data $D \in \mathcal{D}$ to the share calculator.

1.2 **Share calculator:** The share calculator executes the following processing.
   1.2.1 **Share calculation:** Record the receipt time $t_1$ of $D$, sent by the data owner. Then calculate shares $s = (s_1, \ldots, s_{n_{SH}})$ of $D$ and send $s_h$ to share holder $h$.
   1.2.2 **Calculation of tag σ :** Choose $R_{MAC}$ randomly ($R_{MAC} \in_R \mathcal{R}_{MAC}$) and calculate a MAC tag $\sigma = u(R_{MAC}, t_1|D)$, where $t_1|D$ denotes the concatenation of $t_1$ and $D$. Then send $t_1$ and $\sigma$ to the verifier via an authenticated channel (cf. item A4 in Sec. 2.3.1).
   1.2.3 **Post-processing:** Record $t_1, R_{MAC}$ and erase $D, \sigma$ from the memory.

1.3 **Verifier** records the receipt time $t_2$ of messages $t_1$ and $\sigma$, sent by the share calculator.

**(2) Data reconstruction phase:**
2.1 **Initiation by the data owner:** The data owner requests the share calculator to send $D$.

2.2 **Data reconstruction:**
   2.2.1 The share calculator collects from the share holders shares $s = (s_1, \ldots, s_{n_{SH}})$ of $D$. If he could collect only less than $t_{SH}$ shares, he announces "abort" of the entire scheme. From the collected shares, he calculates a data $D'$ (which is supposed to equal $D$) and sends it to the data owner.
   2.2.2 The data owner sends $t_1$ and $D''$ to the end user. If the data owner is honest, $D'' = D'$.

2.3 **Integrity check by the end user:**
   2.3.1 The end user sends $D''$ and $t_1$ to the share calculator.
   2.3.2 The share calculator calculates a MAC tag $\sigma'' = u(R_{MAC}, t_1|D'')$. He then sends $\sigma''$ and $t_1$ to the verifier.
   2.3.3 The verifier verifies the integrity of $D''$, by searching his memory for MAC tag $\sigma$ registered with time $t_1$ and by checking if it satisfies $\sigma'' = \sigma$ and $t_1 \leq t_2$. If the check was successful, he sends "success" to the data owner and the end user; and otherwise send "fail".

### 2.4 Security of our scheme

We state and then prove the security of our scheme.

**Theorem 1:** *Our verification scheme defined in Sec. 2.3 satisfies security criteria SC1, SC2, and SC3, specified in Sec. 2.2.2.*

*Proof:*

SC1: This corresponds to the case where the data owner chooses an incorrect data $D''(\neq D')$ registered with time $t_1$ without knowing $R_{MAC}$, and sends $D''$, instead of $D$, to the end user in step 2.2.2. An honest end user can detect this alteration of $D$ in step 2.3.3, except with a probability no larger than $2^{-k} \log_2 |\mathcal{D}|$, due to the almost universality$_2$ (cf. Lemma 1) of function $u$.

SC2: This corresponds to the case where, after the scheme is finished, the end user chooses an incorrect data $D''(\neq D')$ without knowing $R_{MAC}$, and claims that the data owner sent $D''$ with time $t_1$ instead of $D'$ in step 2.2.2. The data owner, if honest, can refute such claim by

i. calculating the MAC tag $\sigma'' = u(R_{MAC}, t_1|D'')$, with the help of the share calculator, and
ii. then demonstrating to the end user, with the help of the verifier, that $\sigma''$ differs from the correct tag $\sigma$ stored by the verifier.

This refutation fails only when $\sigma''$ equals $\sigma$ accidentally, which occurs with a probability no larger than $2^{-k} \log_2 |\mathcal{D}|$, again due to the almost universality$_2$ of function $u$.

SC3: This is because the verifier is correlated only through the MAC tag $\sigma \in \{0,1\}^k$.

## (1) Data registration phase

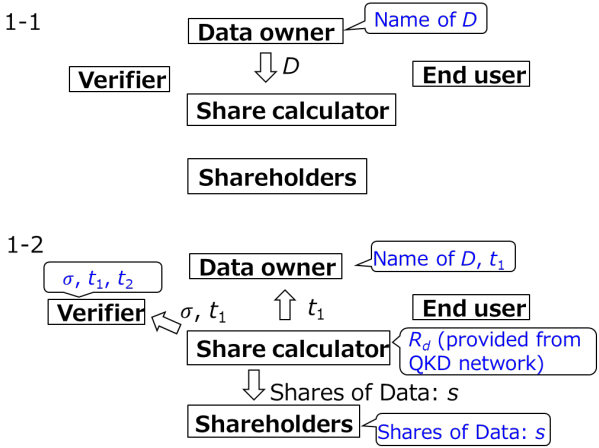

## (2) Data reconstruction and verification phase

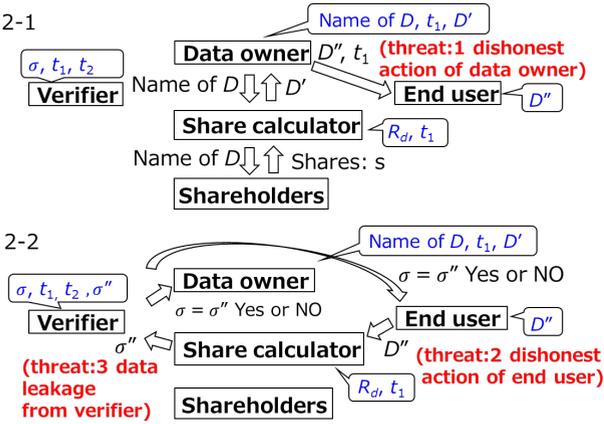

FIGURE 2. Conceptual view of third-party verification scheme. Long-term means elapsed time between (1) data registration phase and (2) data reconstruction and verification phase. The end user is not always fixed at data registration phase.

### 2.5 Optional: Simplified scheme achieving a weaker security (computational security)

So far, we have restricted ourselves with a scheme achieving information theoretical security. In this subsection, we consider a case with the computational security, a weaker notion of security, and show that it admits a simplification of our scheme.

The basic idea is simply to replace the almost universal$_2$ hash function u, used in the above scheme, with a computationally collision resistant hash function [14], instantiated, e.g., by SHA-512.

More precisely, we prepare an optional mode in which the data owner and the end user themselves calculate hash values $h(t_1|D)$ using the SHA-512 function h. In this option, they need not use the share calculator. The data owner calculates $h(t_1|D)$ and sends it to the verifier. The end user, after receiving $D'''$, calculates $h(t_1|D'')$ and sends it to the verifier. The verifier checks if $h(t_1|D'') = h(t_1|D)$ and $t_1 \leq t_2$ hold and informs the end user of the result.

While this option can omit the share calculator, the data integrity (SC1 and SC2) must be somewhat mitigated; it can no longer be guaranteed in the sense of ITS, but only in the sense of computational security. That is, the success probability of an attack in SC1 or SC2 can no longer be bounded by $2^{-k} \log_2 |\mathcal{D}|$, but can only be shown to be negligible (with respect to the security parameter k) against a probabilistic polynomial-time (PPT) malicious user [6].

On the other hand, the secrecy (SC3) still holds, as long as we set the MAC tag length to be smaller than k (see the proof of Theorem 1).

### 2.6 Advantage of our schemes (for both information theoretical security and computational security)

If there is a conflict during the authentication process, the verifier enables to judge whether the data owner or the end user is correct. As mentioned above, no commitment scheme for data integrity application can achieve instant ITS binding and hiding [19]. Our third-party verification scheme is based on the use of a trusted third-party [24]. The novelty of our scheme is to introduce the share calculator to guarantee an integrity of the data. The trusted assumptions on processing hardware are practical and have often been introduced in secure multiparty computation studies [20, 25]. In our case, we require that share calculations are performed secretly and the memory used to calculate MAC tags is long-term secure but small. We think these assumptions are acceptable for practical use.

### III. SPSS scheme and share renewal process

In the previous section, we introduced the third-party verification scheme which can be combined with an arbitrary secret sharing (SS) scheme, SS. In this section, we choose the underlying SS scheme, SS, to be the single-password-authenticated secret sharing (SPSS) given in our previous paper [5], and discuss details of the combined scheme. Then we present its demonstration on an actual QKD network testbed, called the Tokyo QKD Network [8] in section 4. Our SPSS scheme [5] achieves ITS data transmission, storage, authentication, and integrity. In particular, the data owner can verify the integrity of the data when he reconstructs the data by himself. By combining this scheme with the third-party verification scheme of the previous section, we can enhance the data integrity guarantee function. In this section, we outline the procedure of the combined scheme. Further details and the security proof of the SPSS scheme are described in the supplemental information in Ref. [5]. We introduce the (3,4)-threshold scheme below.

(1) Registration phase
(1-1) The data owner sends data D and password P to the share calculator. The share calculator informs received time $t_1$ to the data owner. For the efficiency of calculation, Mersenne primes should be used in the following calculation. Of course, other primes can be applied. To better understanding, we show an example of calculation with Mersenne prime. Since each calculation in the finite field with prime order $q = 2^m - 1$ can deal with only blocks of length at most $m - 1$ bits, secret data $D$, which has generally a much longer length, needs to be divided into pieces of $(m - 1)$-bit block, e.g. $l$ pieces; $D = D_l | D_{l-1} | \cdots | D_1$. The data owner sets a $(m - 1)$-bit password $P$, which should have sufficient entropy against the on-line dictionary attack, then computes a message authentication code, MAC tag= $D_l P^l + D_{l-1} P^{l-1} + \cdots + D_1 P$, which is denoted as $D^{l+1}$, and finally adds it to the data for later purpose of message authentication by the data owner.

(1-2) For each data block, data shares $f_{D_i}(1), f_{D_i}(2), f_{D_i}(3), f_{D_i}(4)$ are created for share holders 1, 2, 3, and 4, respectively, by using polynomial $f_{D_i}$ of degree at most 2, where $i = 1, \ldots, l + 1$. Password shares $f_P(1), f_P(2), f_P(3), f_P(4)$ are created by using polynomial $f_P$ of degree at most 1.

(1-3) They are then sent to the corresponding share holders from the share calculator.

(1-4) Each share holder stores the set of shares.

(1-5) Simultaneously, the share calculator computes the other MAC tag of $t_1 | D$ by using random number $R_{MAC}$ for verification, and sends $t_1$ and the MAC tag to the verifier as mentioned in section 2. $R_{MAC}$ and $t_1$ are stored in the share calculator. Here, we used the Toeplitz matrix-multiplication [26] as the almost universal$_2$ hash function $u$ (satisfying the property of Lemma 1) for calculating the MAC tag.

Overall process of (1-1) to (1-5) are shown in Fig. 3 (1).

(2) Pre-computation and communication phase

(2-1) Each share holder generates a random number, denoted as $R_j$ for the $j$-th storage server, and makes its shares $f_{Rj}(1), f_{Rj}(2), f_{Rj}(3), f_{Rj}(4)$ by using polynomial $f_{Rj}$ of degree at most 1. Furthermore each server generates shares of the "0" $f_{0j}(1), f_{0j}(2), f_{0j}(3), f_{0j}(4)$ by using polynomial $f_{0j}$ of degree at most 2, such that $f_{0j}(0) = 0$ should hold so as to keep confidentiality of the share in data reconstruction phase without changing the value of the data share.

(2-2) The share holders send these shares to each other.

(2-3) Each share holder receives three shares of three random numbers and three shares of the "0," and stores them together with the ones produced by itself.

For ITS, the above procedure has to be iterated $l + 1$ times before each data reconstruction of l blocks secret data. That is, $j$-th share holder has to keep $l + 1$ sets of $(f_{R1}(j), f_{R2}(j), f_{R3}(j), f_{R4}(j), f_{01}(j), f_{02}(j), f_{03}(j), f_{04}(j))$.

These processes are shown in Fig. 3(2-1-2). And share renewal process described in appendix are summarized in Fig. 3 (2-3)-(2-6).

(3) Data reconstruction phase

We assume $P'$ be the password in the data owner's memory.

(3-1) The data owner sends $P'$ to the share calculator. And the share calculator chooses three share holders among the four to recover data. We may assume that they are share holder 1, 2, and 3 without loss of generality, denoting them as a set $L = \{1, 2, 3\}$. The share calculator generates shares of $P'$, $f_{P'}(1), f_{P'}(2), f_{P'}(3)$ by using polynomial $f_{P'}$ of degree at most 1. Each set $(L, f_{P'}(j))$ is sent to each corresponding share holder (request). If $|L| \neq 3$, the request is rejected regarding it as an improper request. Otherwise, for each data block, each server, say $j$-th one, computes $R = f_{R1}(j) + f_{R2}(j) + f_{R3}(j)$, $Z = f_{01}(j) + f_{02}(j) + f_{03}(j)$ and $F_{ji} = (f_P(j) - f_{P'}(j))R + Z + f_{D_i}(j)$.

The $F_{ji}$ ($i = 1, \ldots, l + 1$) are then sent to the share calculator (response). Here note that $R$ and $Z$ should be discarded at each request-response for ITS.

(3-2) For each data block, the share calculator computes polynomial $F_i(x)$ of degree 2 that satisfies $F_i(j) = F_{ji}$ for all $j$. $F_i(0)$ is the reconstructed block.

(3-3) The share calculator computes the MAC tag from $F_1(0), \ldots, F_l(0)$ by using the password. If $F_{l+1}(0)$ is equal to calculated MAC tag, the share calculator determines that the stored data has been successfully reconstructed and sends the data to the owner. If necessary, the data owner or the end user can check the data integrity as mentioned in section 2.

These processes are shown in Fig.3 (3-1)-(3-3).

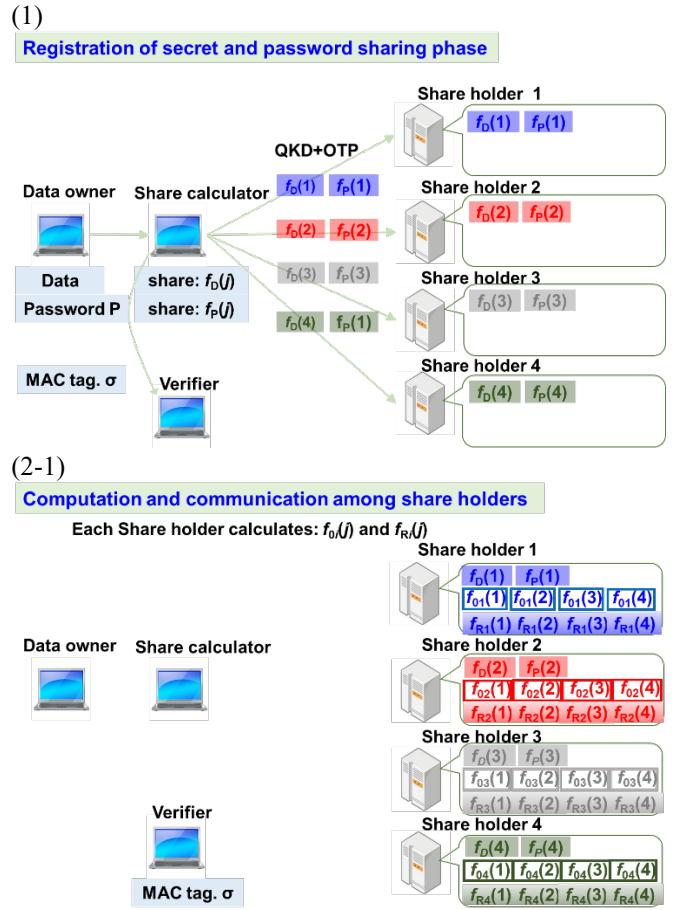

(2-2)
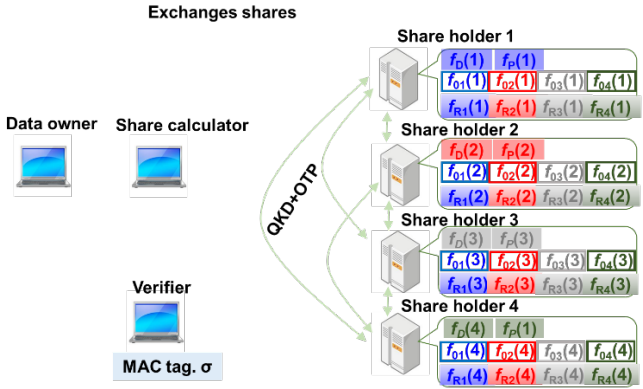

(2-5 optional)
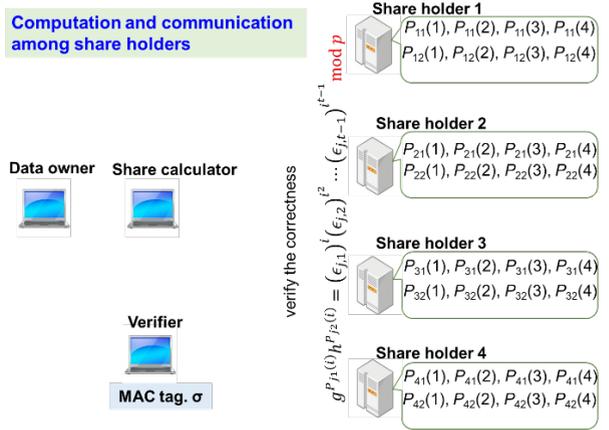

(2-3 optional)
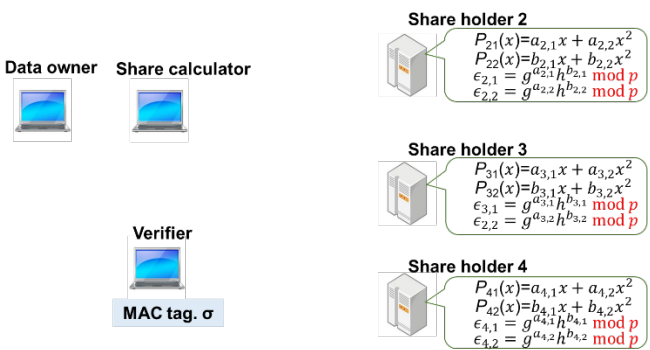

(2-6 optional)
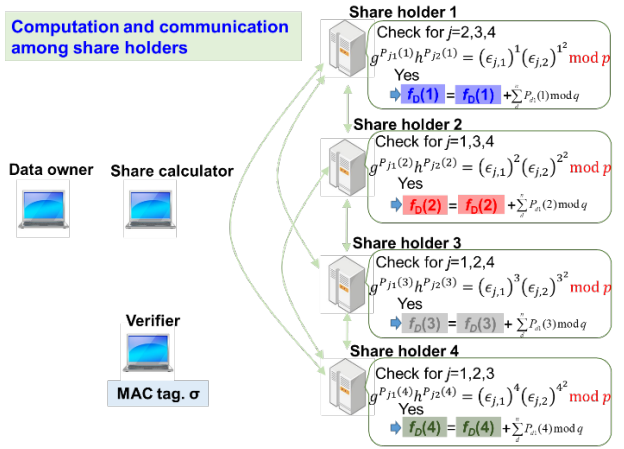

(2-4 optional)
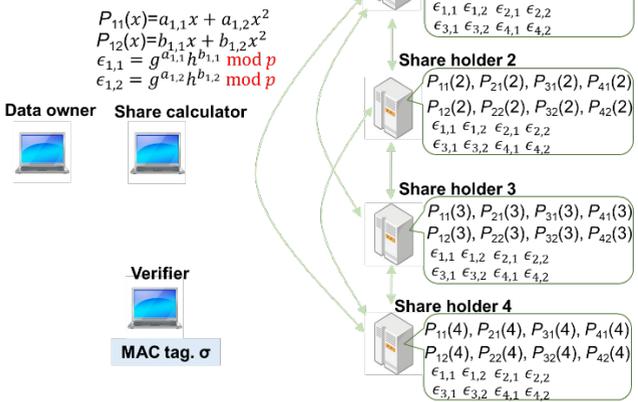

(3-1)
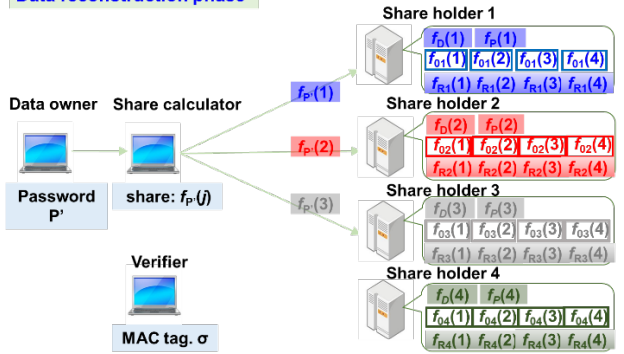

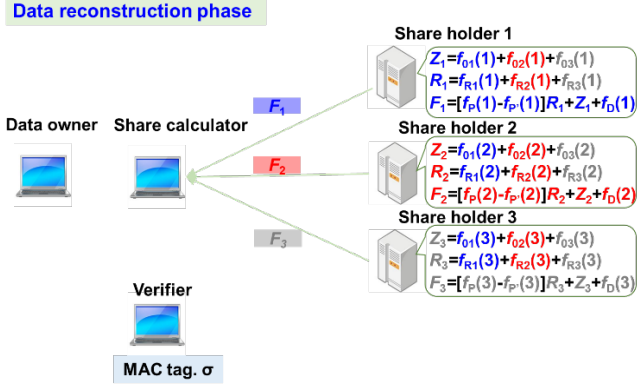

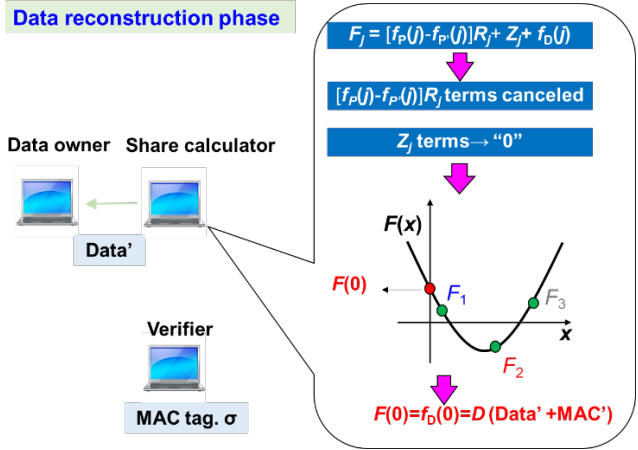

**FIGURE 3.** Schematic diagram of distributed storage with password authenticated secret sharing and share renewal scheme.

## IV. Experimental setup and demonstration on the QKD Network

The third-party verification scheme described in sections 2 and 3 is experimentally demonstrated in a QKD network testbed. As mentioned above, the QKD network can provide not only secure communication lines but also random numbers generated from physical random number generators, because these devices are inside QKD systems. The random numbers generated by intrinsically non-deterministic physical processes are useful for various crypt applications.

The verifiable share renewal function is useful to realize long-term security. Several verification methods have been reported [27-29]. We use Pedersen's protocol [28] for a verification of the share renewal among the share holders. In our scheme, the share holders renew shares by adding shares of "0" with verifying whether their partial information is correct or not. The detailed process is described in the Appendix. This scheme enables a verification of the share renewal with ITS hiding but with computationally secure binding. This verification scheme relies on the hardness of discrete logarithm problem. Therefore, if an eavesdropper and/or an adversary has a quantum computer, the data integrity in this scheme (binding or often called correctness of the data) can be compromised. However, this scheme would still be useful to protect malicious insiders who do not have quantum computers. In fact, the share renewal can be carried out before the number of the compromised share holders exceeds the threshold. Furthermore, outsiders cannot get information about share renewal, because transmission lines are encrypted by OTP. Note that even if this process is eavesdropped, the information of the secret data is not leaked. This share renewal function is the optional countermeasure against malicious classical cyber attack.

*4.1 QKD network structure*

The structure of our system is shown Fig. 4. The QKD network [8] works as a secure key supply infrastructure. Secret sharing or other services are installed in this QKD network. The data owner, the share calculator, share holders, the verifier, and the end user communicate through OTP encrypted communication lines in which secure key are provided from the QKD network. The share calculator also requests random number to the key supply agent (KSA) of the QKD network to calculate shares or MAC tags. Once supplied with the keys or random numbers, the key data in the QKD network are erased and the responsibility of key management moves to application users. Generated keys in each QKD link are pushed up to servers, called key management agents (KMAs). Each KMA is set in a physically protected place, referred to as "a trusted node". A KSA is integrated to the KMA. The KSAs supply users the keys. A key management server (KMS) gathers link information and instructs KMAs to execute key relay according to request from the application layer.

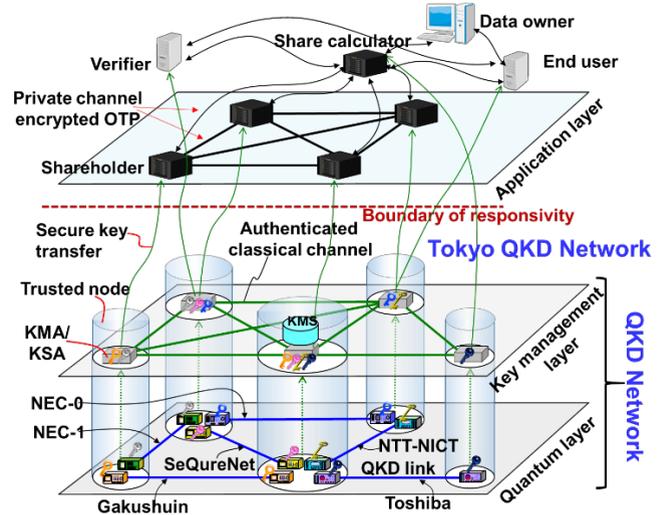

**FIGURE 4.** Schematic view of the layer structure of QKD Network and our third-party verification system.

TABLE I
SPECIFICATION OF QKD LINKS

|  | Protocol | Transmission | |
|---|---|---|---|
|  |  | Length (km) | Loss (dB) |
| NEC-0 | BB84 with decoy | 50 (Spooled fiber NICT premise) | 10 |
| NEC-1 | BB84 with decoy | 22 (field installed 95% areal line) | 13 |
| Toshiba | BB84 with decoy | 45 (field installed 50% areal line) | 14.5 |
| NTT-NICT | DPS-QKD | 90 (field installed 50% areal line) | 28.6-30.1 |
| Gakushuin | CV-QKD | 2 (NICT premise) | 2 |
| SeQureNet | CV-QKD | 2 (NICT premise) | 2 |

The Tokyo QKD Network consists of five nodes (called Koganei-1-4 and Ohtemachi-1) connected by six QKD links. The QKD links consist of the QKD systems provided by NEC [30], Toshiba [31], NTT-NICT [32], Gakushuin [33], and SeQureNet [34]. Specification of each QKD link are listed in Table 1 [5,13]. The key relays are carried out in the key management layer of the Tokyo QKD Network with OTP. Each communication line in the application layer is also encrypted by an OTP manner and authenticated with MAC tag based on Wegman-Carter [15] protocol by using the key from the QKD network. Therefore, each player uses ITS communication. Moreover, high quality physical random number is provided to the share calculator to calculate shares and MAC tags from the KSA of the QKD network.

In the application layer, servers of the data owner, the share calculator, share holders, the verifier, and the end user are set. In this experiment, the data owner is set in Ohtemachi-1. Each share holder (1-4) is located in Koganei-1-4. The verifier and the share calculator are set in Koganei-1 and Koganei-2 respectively. The terminal of the end user is established in Koganei-3.

*4.2 Experimental results*

A prime number $q$ is used in calculation of shares with Galois field. A condition to carry out the share renewal, $q$ must be a divisor of $p$-1. We selected $p$ and $q$ from data sets in [35]. From the viewpoint of fast computation, $q$ should be selected from Mersenne primes. However, it is not so easy to find a prime to meet conditions. Therefore, we selected $p$ and $q$ from data sets in [35].

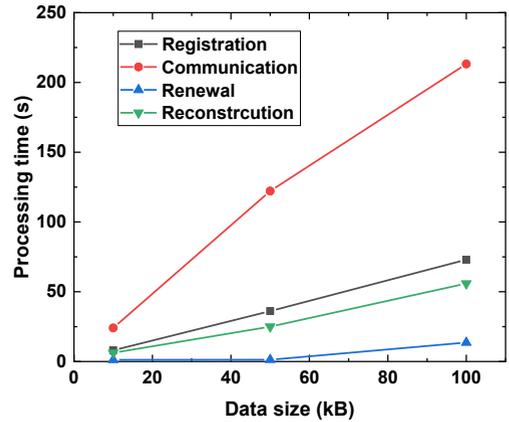

**FIGURE 5.** Processing time of the registration phase (Registration), the communication and communication among share holders phase (Communication), the share renewal phase (Renewal), and the data reconstruction phase (Reconstruction) as functions of data size.

When we carried out this protocol, threshold was set (3,4) shown in Fig.3. The experimental results are shown in Fig. 5. These results indicate practical processing time including data transfer time. Compared with previous results, about three-fold processing time are necessary. It is because we did not use Mersenne prime in calculation of shares. To improve throughput, we shall find Mersenne primes which meet conditions mentioned above.

## V. Conclusion

We propose and demonstrate third-party verification with information theoretical security in a distributed storage system built on the QKD network. We demonstrated, for the first time to our best knowledge, a distributed storage system with information theoretically secure data transmission, storage, authentication, and data integrity with the third-party verification in a real metropolitan area network. By establishing the trusted share calculator and the verifier, falsifying data by the data owner or the end user become extremely difficult, and accidental data leakage from the verifier can be also protected with information theoretical security. We use the universal$_2$ hash function to calculate MAC tags, therefore, calculation load can be extenuated compared with SHA families. It may imply that our scheme is suitable to guarantee integrity of big data. Moreover, we add share renewal function on our previous system "singe-password-authenticated secret sharing (SPSS) system" which enables ITS authentication, data transfer, data storage, and data reconstruction for resisting to against classical cyber attack. In verification scheme, we use universal$_2$ hash function, however, we also developed another option with strongly universal$_2$ hash function. This option enables to prevent accidental data leakage more efficiently, though more random numbers are necessary [6,14]. Our proposed scheme will make a significant contribution to enhancing function of the long-term secure distributed storage system. Our scheme consists of five constituent members (data owner, share calculator, share holder, end user, and verifier) because our scheme includes time-stamp and verification functions. On the other hand, there exists a simpler ITS commitment scheme [24] and its realization with the QKD network is an interesting future work.

Another important future direction is an improvement of the share renewal process. We demonstrated the optional-simple share renewal function based on Pedersen's verifiable secret sharing scheme against malicious insiders or classical cyber attack. The security of this share renewal is based on computational complexity, therefore, it is efficient only against malicious attackers who do not have quantum computers and need a certain time to crack the share holders. It future, it is desirable to realize long-term integrity based on a novel ITS share renewal function, e.g. using a trusted calculator.

## ACKNOWLEDGMENT

This work was partly supported by Council for Science, Technology and Innovation (CSTI), Cross-ministerial Strategic Innovation Promotion Program (SIP), "Photonics and Quantum Technology for Society 5.0" (Funding agency: QST), Ministry of Internal Affairs and Communications "ICT Priority Technology Research and Development Project (JPMI00316)" and JSPS KAKENHI Grant Number JP18H05237. The authors would like to thank NEC, Toshiba, Gakushuin University, and NTT for their contribution to an operation of the Tokyo QKD Network.

## Appendix

The detailed process of optional share renewal shown in Fig.3 (2-3)-(2-6) is listed as follows;

1. Each $i$'th shareholder $i \in [1 \dots n]$ randomly picks $t$-1 numbers from the finite field. These numbers define a polynomial $P_i(x)$ of degree $t$-1 whose free coefficient is zero ($P_i(0) = 0$). $P_{i1}(X) = a_{i,1}x + a_{i,2}x^2 + \cdots + a_{i,t-1}x^{t-1} \mod q$, $P_{i2}(X) = b_{i,1}x + b_{i,2}x^2 + \cdots + b_{i,t-1}x^{t-1} \mod q$. And each $i$'th share holder selects common security parameters $g$ and $h$, and computes

$$\epsilon_{i,j} = g^{a_{i,j}} h^{b_{i,j}} \mod p (j = 1, \dots, t-1). (g = h^x)$$

Shown in Fig.3 (2-3).

2. Each $i$'th shareholder computes the shares $P_{i1}(1), \cdots P_{i1}(n)$, and $P_{i2}(1), \cdots P_{i2}(n)$. And $\left(E(P_{i1}(c), P_{i2}(c))\right)$ is encrypted with one time pad using key from QKD.3. Each $i$'th shareholder publishes the message: $(i, \epsilon_{i,j})_{j \in \{1 \dots t-1\}}$, $E(P_{i1}(c), P_{i2}(c))_{c \in \{1 \dots n\} \setminus i})$ and a signature each other. Shown in Fig.3 (2-4).

4. For all such messages (from other servers) that the $i$'th shareholder receives, he/she verifies the correctness of shares using the same equation as

$$g^{P_{j1}(i)} h^{P_{j2}(i)} = (\epsilon_{j,1})^i (\epsilon_{j,2})^{i^2} \cdots (\epsilon_{j,t-1})^{i^{t-1}} \mod p$$

Shown in Fig.3 (2-5).

5. If the $i$'th shareholder finds all the messages sent by other shareholders to be correct (e.g., correct signature, correct time period, etc.), and all the above equations hold, then it broadcasts a signed acceptance message announcing that all the checks are successful.

6. If all shareholders sent such acceptance messages, the $i$'th shareholder updates his/her own share by $h(i) = f(i) + \sum_{d=1}^{n} P_{d1}(i) + P_{d2}(i) \mod q$.

Shown Fig. 3 (2-6).

7. If in step 5, the $i$'th shareholder finds any irregularities in the behavior of any shareholder during step 4, then he/she broadcasts a signed accusation against the misbehaving shareholder.

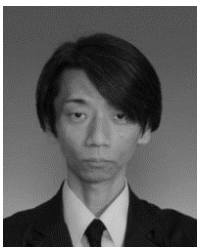

**MIKIO FUJIWARA** received the B.S. and M.S. degrees in electrical engineering and the Ph.D. degree in physics from Nagoya University, Nagoya, Japan, in 1990, 1992, and 2002, respectively. He has been involved R&D activities at NICT (previous name CRL, Ministry of Posts and Telecommunications of Japan) since 1992. His main research interests include quantum key distribution, and QKD network application.

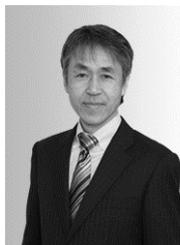

**MASAHIDE SASAKI** received the B.S., M.S., and Ph.D. degrees in physics from Tohoku University, Japan, in 1986, 1988 and 1992, respectively. During 1992-1996, he worked on the development of semiconductor devices in Nippon-Kokan Company (currently JFE Holdings). In 1996, he joined the Communications Research Laboratory, Ministry of Posts and Telecommunications (since 2004, NICT). He has been working on quantum optics, quantum communication and quantum cryptography. He is presently Distinguished Researcher of Advanced ICT Research Institute, and NICT Fellow.

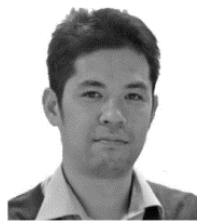

**RYO NOJIMA** is Director of Security Fundamentals Laboratory, Network Security Research Institute, NICT. His main research interests include privacy, cryptography, and information security, especially privacy preserving protocols.

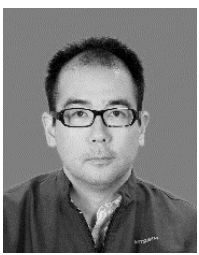

**TOYOHIRO TSURUMARU** was born in Japan in 1973. He received the B.S. degree from the Faculty of Science, University of Tokyo, Japan in 1996, and the M.S. and Ph.D. degrees in physics from the Graduate School of Science, University of Tokyo, Japan in 1998 and 2001, respectively. Then he joined Mitsubishi Electric Corporation in 2001. His research interests include theoretical aspects of quantum cryptography, as well as modern cryptography.

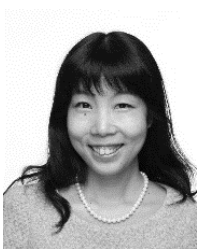

**SHIHO MORIAI** received the B.E. degree from Kyoto University in 1993 and Ph.D. from the University of Tokyo in 2003. She is Managing Director of Strategic Planning Department, and Executive Researcher of Cybersecurity Research Institute, NICT. Her research interests include information security and data privacy.

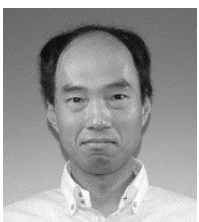

**MASAHIRO TAKEOKA** received the Ph.D. degree in electrical engineering from Keio University, Kanagawa, Japan, in 2001. He is a director of Quantum ICT Advanced Development Center in NICT. His current research interests are in quantum cryptography, quantum information theory, and optical quantum information processing.